\documentclass{elsart}
\usepackage{epsfig} \usepackage{amsmath}
\usepackage{graphicx}%
\usepackage{amsfonts}%
\usepackage{amssymb}
\begin{document}
\begin{frontmatter}
\title{A stochastic Trotter integration scheme for dissipative particle dynamics}
\author[FIS]{M. Serrano\corauthref{cor1}}
\author[CCS]{G. De Fabritiis}
\author[FIS]{P. Espa\~{n}ol}
\author[CCS]{P.V. Coveney}
\address[FIS]{Departamento de F\'{\i}sica Fundamental, UNED\\
Senda del Rey 9, 28040 Madrid, Spain}
\address[CCS]{Centre for Computational Science, Department of Chemistry,\\
 University College London, 20 Gordon street, WC1H 0AJ, London, UK }
 \corauth[cor1]{Corresponding author. E-mail address: mserrano@fisfun.uned.es}
\begin{abstract}
  In this article we show in detail the derivation of an integration scheme
  for the dissipative particle dynamic model (DPD) using the stochastic
  Trotter formula \cite{SDETrotter05}. We explain some subtleties due to the
  stochastic character of the equations and exploit analyticity in some
  interesting parts of the dynamics.  The DPD-Trotter integrator demonstrates
  the inexistence of spurious spatial correlations in the radial distribution
  function for an ideal gas equation of state. We also compare our numerical
  integrator to other available DPD integration schemes.
\end{abstract}
\end{frontmatter}
\section{Introduction}
Mesoscopic models require the use of stochastic differential equations (SDEs)
to include the effects of thermal fluctuations so the selection of an
efficient stochastic integration scheme is crucial to simulate correctly these
systems.  In this article we focus on dissipative particle dynamics (DPD)
\cite{hoog92,espanol95} as one of the most simple and widely used model.
Although nowadays the conventional DPD model has a good theoretical basis, in
the past few years there has been quite a controversy reported in literature
about practical aspects of the simulations: the appearance of spurious effects
related to time discretization, in concrete, the unphysical systematic drift
of the temperature from the value predicted by the fluctuation-dissipation
theorem and uncontrolled spatial correlations among particles. This is the
reason of the increasing interest in developing good integrator methods for
the DPD model.  Several authors \cite{groot97,pagonabarraga98,besold00} have
considered improvements to the basic stochastic Euler scheme through the use
of solvers that have been successfully employed for \emph{deterministic}
dynamical systems in molecular dynamics (MD) simulations \cite{tuckerman92}
such as the velocity Verlet algorithm (DPD-VV \cite{groot97}).

Recently in \cite{SDETrotter05} we investigated the applicability of the
Trotter formula (widely used in molecular simulations) to a general SDEs and
discuss the optimum way to split the dissipative-stochastic generators. It
resulted that the Trotter formula cannot be applied without considering the
special stochastic character of the equations. In general, different variables
depending on the same noise should not be split. In the DPD equations this
does not happen which allowed us to write a integration scheme.  In
\cite{SDETrotter05} we concluded that, considering the accuracy of the
equilibrium temperature and the computational cost, DPD-Trotter is among the
best integrators for the DPD equations.

In this article we explain in detail how to apply the stochastic Trotter
formula to the particular case of DPD.  The aim is to furnish a non-trivial
example to be used as a reference when one wishes to derive new integration
schemes based on the stochastic Trotter formula for a general set of SDEs. We
also test the behavior of the radial distribution function in the DPD model.
For an ideal gas equation of state we find the DPD-Trotter
scheme presents no spatial correlations at any scale.
\section {A Trotter integration scheme for dissipative particle dynamics}
\label{section_dpd}
The DPD model consists of a set of $N$ particles moving in continuous space.
Each particle $k$ is defined by its position $\mathbf{r}_{k}$ and its momentum
$\mathbf{p}_{k}$ and mass $m$. The dynamics is specified by a set of Langevin
equations very similar to the molecular dynamics equations, but where in
addition to the conservative forces there are dissipative and fluctuating
forces as well
\begin{equation}
\begin{array}{l}
d\mathbf{{r}}_{k} =\frac{\mathbf{p}_{k}}{m} dt,\\
d\mathbf{{p}}_{k} =\sum_{l\neq k}^{N} \mathbf{e}_{kl}\left [  \left ( a_{kl} F_c(r_{kl})
 - \frac{\gamma}{m} \omega^2(r_{kl})(\mathbf{e}_{kl}\cdot
\mathbf{p}_{kl})\right ) dt + \sqrt{2\gamma k_B T_0} \omega(r_{kl})dW^t_{kl}\right ],
\end{array}
\label{dpd}
\end{equation}
where $F_c(r)$ is the conservative pair interaction force weighted by positive
and symmetric parameter $a_{kl}$, $\mathbf{r}_{kl}={\bf r}_k - {\bf r}_l$ is
the distance between the particle $k$ and particle $l$, ${r}_{kl}$ its length
and ${\bf e}_{kl}={\bf r}_{kl}/r_{kl}$.  The weight function $\omega$ usually
has a finite range $r_c$.  A typical selection is $\omega(r)=1-r/r_{c}$ for
$r<r_{c}$ and $\omega(r)=0$ for $r\geq r_{c}$.  The conservative force is
usually chosen to be of the form ${ F}_c(r_{kl})=w(r_{kl})$. This system has a
well defined Gibbsian equilibrium state at a temperature $T_0$. Because the
stochastic term in conventional DPD does not depend on the momenta, note the
It\^{o} or Stratonovich interpretation are exactly equivalent (additive noise) and
we can apply the standard rules of ordinary calculus formally treating $dW$ as
$''dt''$.

The global state is ${\bf x}=({\bf r}_1,{...},{\bf r}_N,{\bf p}_1,{...},{\bf
  p}_N)$ and the SDEs (\ref{dpd}) can be expressed as $d{\bf
  x}=\mathcal{L}[{\bf x}] dt$ with formal solution ${\bf
  x}(t)={\mathcal T} e^{\mathcal{L}t}[{\bf x}]({\bf x}_0)$ where ${\mathcal
  T}$ is the time-ordered operator (see \cite{ricci03}). The global
time generator is divided in two operators generating ``orthogonal dynamics''
$\mathcal{L}=\mathcal{L}_r+\mathcal{L}_p$, where 
$\mathcal{L}_r=\sum_k
\mathcal{L}^k_r$ and $\mathcal{L}_p=\sum_{k,l> k}\mathcal{L}^{kl}_p$. Because in
the DPD model the forces between interacting particles $k$ and $l$ satisfy
action-reaction (Newton's third law), the momentum is locally (and totally)
conserved. For each partial dynamics, the generator can be subdivided in
components $\mathcal{L}^k_r =\sum_\mu \mathcal{L}_{\mathbf{r}_{k}^\mu}$ and
$\mathcal{L}^{kl}_p = \sum_\mu\mathcal{L}_{\mathbf{p}_{kl}^\mu}$ with
\begin{eqnarray}
\mathcal{L}_{\mathbf{r}_{k}^\mu}&=& \frac{\mathbf{p}_{k}^\mu}{m}
\partial_{{\bf r}_k^\mu}, \quad \quad  \quad \quad 
\mathcal{L}_{\mathbf{p}_{kl}^\mu} =\mathcal{D}_{\mathbf{p}_{kl}^\mu}
+\mathcal{S}_{\mathbf{p}_{kl}^\mu}(t), \nonumber\\
\mathcal{D}_{\mathbf{p}_{kl}^\mu}&=&\left [ a_{kl} F_c(r_{kl}) 
 -\frac{\gamma}{m} \omega _{D}(r_{kl})(\mathbf{e}_{kl}\cdot 
\mathbf{p}_{kl})\right ] \mathbf{e}^\mu_{kl}
(\partial_{{\bf p}_k^\mu}-\partial_{{\bf p}_l^\mu}),\nonumber\\
\mathcal{S}_{\mathbf{p}_{kl}^\mu}(t)&=&f(t)\sqrt{2\gamma k_B T_0} 
\omega(r_{kl})\mathbf{e}^\mu_{kl} (\partial_{{\bf p}_k^\mu}-\partial_{{\bf p}_l^\mu}).
\end{eqnarray}  
Note that the momentum operator has two contributions, the deterministic and
the stochastic which is an explicit function of time with
$f(t)=dW^t_{kl}/dt$.

The formal solution of our system ${\bf x}(t)={\mathcal T}
e^{\mathcal{L}t}[{\bf x}]({\bf x}_0)$ corresponds to a continuous time
evolution. In order to devise any integrator scheme we must discretize the
continuous time in finite steps.  The continuum time propagator can be
approximated by discrete time steps of size $\Delta t=t/P$, recursively
applying $P$ times the exponential operator $e^{\mathcal{L}t} \equiv \left
  (e^{\mathcal{L}t/P}\right)^{P}\approx e^{\mathcal{L}\Delta t}\cdots
e^{\mathcal{L}\Delta t}$.  Note that when the generator depends explicitly on
time $\mathcal{L}(t)\equiv \mathcal{L}^t$, the time-ordered exponential is
relevant and the recursively nested exponentials become ${\mathcal T}
e^{\mathcal{L}^tt} \approx e^{\mathcal{L}^{t+P\Delta t}\Delta t} \cdots
e^{\mathcal{L}^{t+2\Delta t}\Delta t} e^{\mathcal{L}^{t+\Delta t}\Delta t}$
(\cite{ricci03}).  At this point we must provide some approximation of the
discrete time propagator of a ``generic'' global dynamics
$e^{\mathcal{L}\Delta t}$. As we have mentioned in the previous paragraph, in
general the generator ${\mathcal L}$ is formed by many generators
$\sum{\mathcal L}_i$ each of them corresponding to a particular dynamics $i$.
The generalized Trotter formula (Strang \cite{strang68}) generates a
straightforward approximation to the time propagator exact up to second order
in time
\begin{equation}
e^{\sum_{i=1}^{M}\mathcal{A}_it}=\left( \prod_{i=M}^{1}e^{\mathcal{A}_{i}\frac{\Delta t}{2}
}\prod_{j=1}^{M}e^{\mathcal{A}_{j}\frac{\Delta t}{2}}\right) ^{P}+O(\Delta
t^{3}).   
\label{dettrotter}
\end{equation}
Because it can be performed in many possible ways, the most important practical
issue to apply formula (\ref{dettrotter}) is the selection of a particular
splitting of the global dynamics.  One reasonable criteria is to keep the
minimum number of generators and exploit analyticity for each of them whenever
possible. For the stochastic equations of DPD, the splitting we propose
consist in $1+\frac{N(N-1)}{2}$ operators: the global $\mathcal{L}_r$ and a
$\mathcal{L}_p^{kl}$ for each pair.

The Baker-Campbell-Haussdorff (BCH) formula reads
\begin{equation}
e^{\mathcal{A}}e^{\mathcal{B}}=e^{\mathcal{A}+\mathcal{B}
+\frac{1}{2}[\mathcal{A},\mathcal{B}]
+\frac{1}{12}[\mathcal{A},[\mathcal{A},\mathcal{B}]]
+\frac{1}{12}[\mathcal{B},[\mathcal{B},\mathcal{A}]]+\cdots}
\end{equation}
so for $[\mathcal{A},\mathcal{B}]=0$ we have the exact formula
$e^{\mathcal{A}+\mathcal{B}}=e^{\mathcal{A}}e^{\mathcal{B}}=e^{\mathcal{B}}e^{\mathcal{A}}$.
The DPD position  generator
$\mathcal{L}_r=\sum_{k,\mu}\mathcal{L}_{r_k^\mu}$ is composed by many 
simple individual generators per particle and components that satisfy
$[\mathcal{L}_{\mathbf{r}_{k}^\mu},\mathcal{L}_{\mathbf{r}_{l}^\nu}]=0$ for
all particles $k,l$ ($k\ne l$) and components
$\mu,\nu$ except $\mu\ne\nu$. Therefore we can use the exact formula
\begin{equation}
e^{\mathcal{L}_r\Delta  t}=e^{\sum_k\mathcal{L}^k_r\Delta  t}\equiv
\prod_{k=1}^{N}\left(\prod_{\mu=1}^{d}e^{\mathcal{L}_{{\mathbf r}_{k}^\mu}
    \Delta t}\right )
\end{equation}
with $d$ the dimensionality. In MD (DPD without dissipative and random forces)
the momentum generator $\mathcal{L}_p=\sum_{k,\mu}\mathcal{L}_{p_{kl}^\mu}$
also satisfies $[\mathcal{L}_{\mathbf{p}_{kp}^\mu},
\mathcal{L}_{\mathbf{p}_{lq}^\nu}]=0$ for all pairs of particles $kp,lq$, with
$k\ne l$, $p\ne q$, $p\ne q$ and components $\mu,\nu, \mu\ne\nu$, such that
the ordering of the individual-component momentum generators is absolutely
irrelevant.  On the contrary in DPD, the forces depend on the other components
of the velocity of the particle and also on other particles velocities and the
operator
$e^{\mathcal{L}_p\Delta  t}=e^{\sum_{k,l>k}\mathcal{L}^{kl}_{p}\Delta  t}$
cannot be globally integrated and has to be approximated in some way.  This is
the reason for the splitting it in $\frac{N(N-1)}{2}$ momentum operators.  Due
to this splitting and formula (\ref{dettrotter}), the DPD scheme is finally
given by the following Trotter integrator
\begin{equation}
{\bf x}(t+\Delta t)=  
\left (\prod_{q=1,r>1}^{N}
e^{\mathcal{L}^{qr}_p\frac{\Delta t}{2}}\right )
\left (\prod_{i=1}^{N}e^{\mathcal{L}^i_r\Delta t} \right )
\left ( \prod_{k=N,l<N}^{1}
e^{\mathcal{L}^{kl}_p\frac{\Delta t}{2}}\right ){\bf x}(t).
\label{dpdscheme}
\end{equation}
The  propagator that corresponds
to the generator $\mathcal{L}_{{\bf r}_k}^\mu$ produces the position update which is
analytically given by
\begin{equation}
e^{\mathcal{L}^k_{r^\mu_k}\Delta t}[{\bf x}]: \quad 
{ r}_k^\mu (t+\Delta t)={ r}_k^\mu(t)+ \frac{{ p}_k^\mu( t)}{m} \Delta t
\label{schemev}
\end{equation}
because the momentum is a constant in this step of the scheme. The
next step is to solve the propagator of the momenta of the interaction pair
$k$, $l$ (corresponding to the generator
$\mathcal{L}_p^{kl}$) independently of the positions.  We
have mentioned before that DPD forces satisfy action-reaction, so for a
particular interacting pair $k,l$ we propose to make a change of variables
from $\mathbf{{p}}_{k}, \mathbf{{p}}_{l}$ to
$\mathbf{{p}}_{k}+\mathbf{{p}}_{l}, \mathbf{{p}}_{kl}=\mathbf{{p}}_{k}-
\mathbf{{p}}_{l}$.  The new system to solve is $d(\mathbf{p}_{k}+
\mathbf{p}_{l})=0$ and $d\mathbf{p}_{kl}=2d\mathbf{p}_{k}$.
Because the positions of the particles are ``frozen" at this step of the
Trotter scheme, the equation for $d\mathbf{{p}}_{kl}$ can be solved more
easily for the projection on the radial direction
$p^e_{kl}=\mathbf{{p}}_{kl}\cdot\mathbf{e}_{kl}$
\begin{equation} 
dp^e_{kl}= A dt - B p^e_{kl}dt + C dW^t_{kl},
\end{equation}
where $A=2a_{kl}F_c(r_{kl})$, $B=2 \gamma/m \omega^2$ and $C=2 \sqrt{2\gamma
  k_B T_0} \omega$. This equation is an Ornstein-Uhlenbeck process with
analytical solution \cite{kloeden92}
\begin{equation}
p^e_{kl}(t)= e^{-B \Delta t}  p^e_{kl}(t_0) 
+A \int_{t_0}^{t}  e^{B (s-t)}ds
+C \int_{t_0}^{t} e^{B(s-t)}dW_s, 
\label{exact}
\end{equation} 
where $\Delta t = t-t_0$, $t_0$ being the initial time.  The solution
of (\ref{exact})  requires the
generation of colored noise based on a numerical scheme itself.
A version of the method to generate coloured noise \cite{SDETrotter05,fox88}
adapted to Eq.(\ref{exact}) results 
\begin{equation}
\Delta {p}^e_{kl} = \left (\mathbf{{p}}_{kl}\cdot{\bf e}_{kl} - \frac{a_{kl}F_c}{\tau}\right)
 \left(e^{-\frac{2\Delta t }{\tau}  }-1 \right ) 
           + \sqrt{
2 k_B T_0 m
\left (1 -  e^{-\frac{4 \Delta t }{\tau} }\right )}\xi^{kl},  
\end{equation}
where $\tau= \gamma/m \omega^2$, $\xi^{kl}=\xi^{lk}$ are normal distributed
with zero mean and variance one ($N(0,1)$) and $\Delta {p}^e_{kl}={p}^e_{kl}(
t) - {p}^e_{kl}( t_0)$.
The propagator $e^{\mathcal{L}^{kl}_p \Delta t}$ for $\mathbf{p}_{k}$ and
$\mathbf{p}_{l}$ gives
\begin{equation}
e^{\mathcal{L}^{kl}_p \Delta t}[{\bf x}]: 
(\mathbf{{p}}_{k}^{t+\Delta t},\mathbf{{p}}_{l}^{t+\Delta t})=
\left (\mathbf{{p}}_{k} (t)+ \frac{\Delta {p}^e_{kl} }{2}{\bf e}_{kl}(t), \quad
 \mathbf{{p}}_{l} (t)- \frac{\Delta {p}^e_{kl}  }{2}  {\bf e}_{kl}(t)\right).
\label{scheme_nc}
\end{equation}
%
So in DPD we can solve the dynamics corresponding to the generator ${\mathcal
  L}^{kl}_{p}$ (globally for all components at the same time) without the need
to go to the scalar operator $\mathcal{L}_{\mathbf{p}_{kl}^\mu}$ corresponding
to the coordinate $\mu$.  In practice the Trotter integration algorithm
(\ref{dpdscheme}) consists of the following steps: for the interaction pairs
$k,l$ update the momentum half timestep according to the propagator
(\ref{scheme_nc}) with a noise $\xi^{kl}$; iterate over particles $k$ updating
the position according to (\ref{schemev}); finally, update pairs $k,l$ in
reverse order again using the propagator (\ref{scheme_nc}) but with new noises
$\xi'_{kl}$.  This algorithm requires the calculation of the pair-list only
once per iteration and has the same complexity as a simple DPD velocity-Verlet
scheme (DPD-VV \cite{groot97}).

\begin{figure}[t]
\begin{center}
\psfig{figure=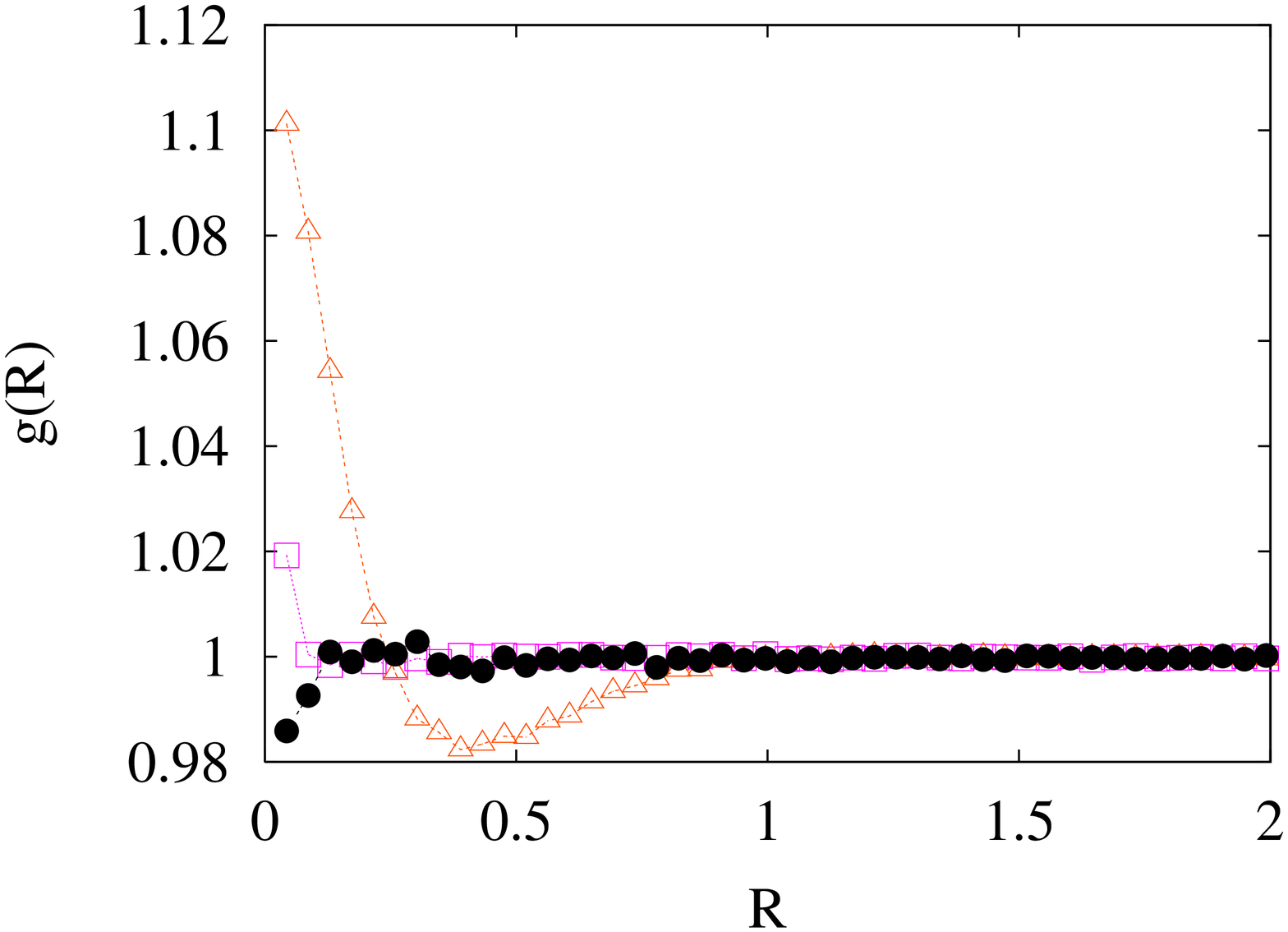,width=5cm, angle=-90}\psfig{figure=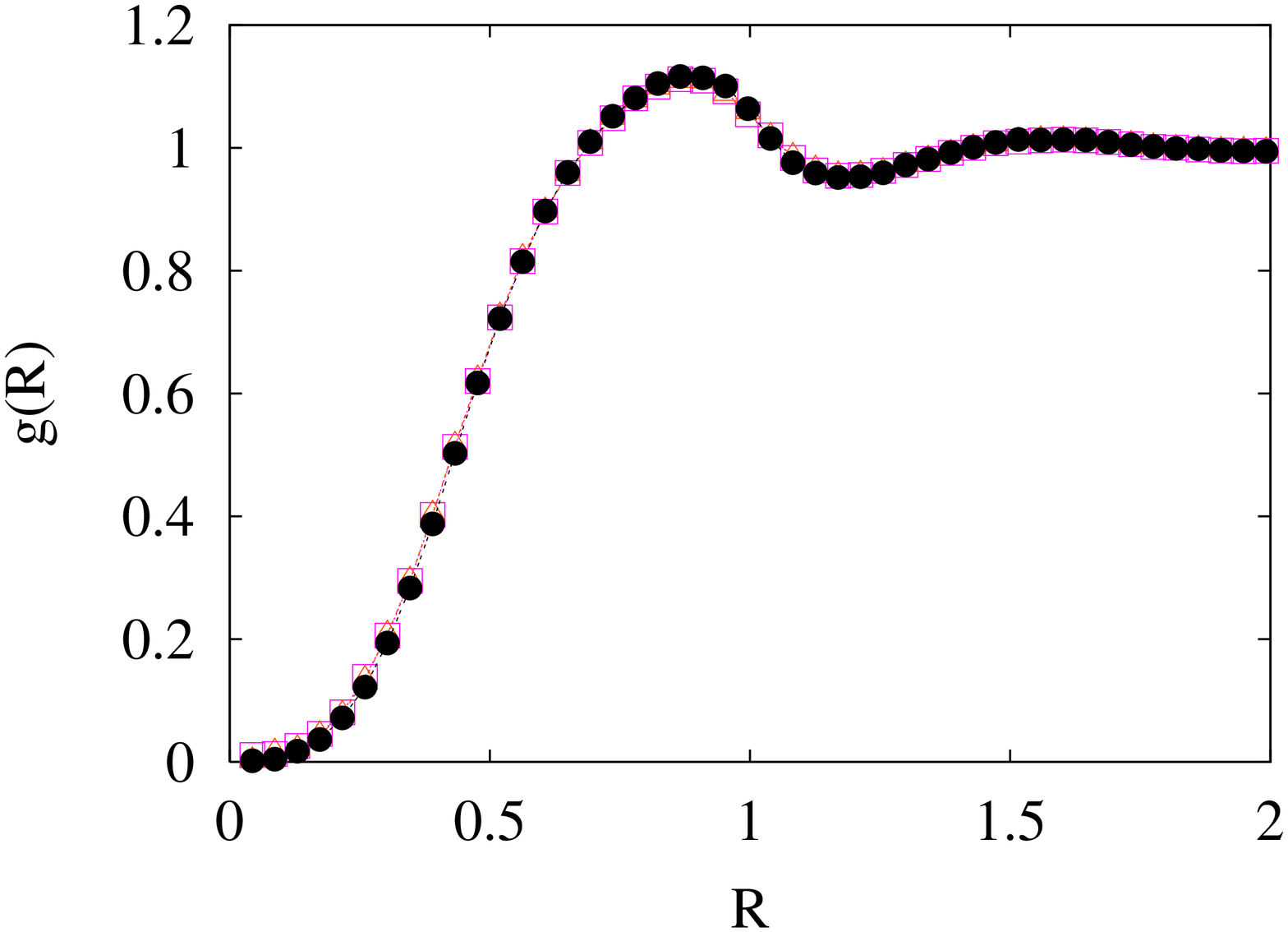,width=5cm, angle=-90}
\end{center}
\caption{Radial distribution function for three integrator methods. Velocity
  Verlet with ($\triangle$) symbols, Shadlow scheme with  ($\Box$)
  symbols and  Trotter DPD with ($\bullet$) symbols. Left figure corresponds to an  ideal gas simulation $a_{kl}=0$. Right figure corresponds to simulations including  conservative forces with  $a_{kl}=25$.}
\label{gdr_fig}  
\end{figure}

We tested in \cite{SDETrotter05} this integration scheme using the open-source
code {\em mydpd} \cite{mydpd} for the equilibrium temperature with $N=4000$
particles, $\gamma=4.5, k_B T_0=1, m=1, r_c=1$ in a three dimensional periodic
box $(L,L,L)$ with $L=10$ with periodic boundary conditions. These settings
give a particle density $\rho=4$. Here, we show in left Fig.\ref{gdr_fig} the
radial distribution function for $a_{kl}=0$ (corresponding to an ideal gas
equation of state) and a time step $\Delta t=0.05$. We compare the results for
three methods: the velocity Verlet (DPD-VV) \cite{groot97}, the Shardlow
scheme \cite{shardlow03} and DPD-Trotter \cite{SDETrotter05}.  We find good
agreement with the theoretical value $1$ for Shardlow and DPD-Trotter
integrators but DPD-VV is notably wrong displaying spurious spatial
correlations at distances less than the finite range $r_c$. In right
Fig.\ref{gdr_fig} we show the radial distribution function for a simulation
with $a_{kl}=25$ and a time step $\Delta t=0.01$. As we see the three methods
perform very similarly.
\section{Conclusions}
The stochastic Trotter formula can be successfully applied to the DPD model and
the procedure to tailor the integrator scheme has been explained in detail.
In the scheme we have also exploited the exact integration of important parts
of the dynamics like the conservation of total momentum of an interacting pair
of particles.  The DPD-Trotter integrator displays correctly the radial
distribution functions for an ideal gas (no conservative forces among
particles) and also for a non ideal gas.  Following this important example and
\cite{SDETrotter05} it should be strathforward to apply the stochastic Trotter
formula to new mesoscopic models and more general SDEs.

{\bf Acknowledgements} M.S.  and P.E. are supported by the Spanish Ministerio
de Educaci\'on y Ciencia project FIS2004-01934 and GDF by the EPSRC
Integrative Biology project GR/S72023. PVC \& MS thank the EPSRC (UK) for
funding RealityGrid under grant number GR/R67699; this project supported MS's
6 month visit to the CCS at UCL during 2005.  
\bibliographystyle{elsart-num}
\bibliography{../gianni}
\end{document}